\documentclass[showpacs, oneside, twocolumn, amsmath, amssymb, prl, nofootinbib, superscriptaddress,aps,showkeys]{revtex4-1}
\usepackage{amsfonts,amsmath,amssymb,array,bm,bbm,cases,color,dcolumn,graphicx,mathrsfs,microtype,subfigure,xcolor}
\usepackage{appendix}
\usepackage{float}
\usepackage[colorlinks,linktocpage,linkcolor=cyan,citecolor=cyan]{hyperref}
\definecolor{lightgray}{gray}{0.9}
\definecolor{Amber}{rgb}{1.0, 0.75, 0.0}
\definecolor{blizzardblue}{rgb}{0.67, 0.9, 0.93}

\begin{document}


\title{Primordial black hole mass functions as a probe of cosmic origin}

\author{Yi-Fu Cai}
\email{yifucai@ustc.edu.cn}
\affiliation{Deep Space Exploration Laboratory/School of Physical Sciences, University of Science and Technology of China, Hefei, Anhui 230026, China}
\affiliation{CAS Key Laboratory for Researches in Galaxies and Cosmology/Department of Astronomy, School of Astronomy and Space Science, University of Science and Technology of China, Hefei, Anhui 230026, China}

\author{Chengfeng Tang}
\affiliation{Deep Space Exploration Laboratory/School of Physical Sciences, University of Science and Technology of China, Hefei, Anhui 230026, China}
\affiliation{CAS Key Laboratory for Researches in Galaxies and Cosmology/Department of Astronomy, School of Astronomy and Space Science, University of Science and Technology of China, Hefei, Anhui 230026, China}

\author{Geyu Mo}
\affiliation{Deep Space Exploration Laboratory/School of Physical Sciences, University of Science and Technology of China, Hefei, Anhui 230026, China}
\affiliation{CAS Key Laboratory for Researches in Galaxies and Cosmology/Department of Astronomy, School of Astronomy and Space Science, University of Science and Technology of China, Hefei, Anhui 230026, China}

\author{Sheng-Feng Yan}
\affiliation{Istituto Nazionale di Fisica Nucleare (INFN), Sezione di Milano, 
20146, Milano, Italy}
\affiliation{DiSAT, Universit\`{a} degli Studi dell'Insubria, 
22100, Como, Italy}

\author{Chao Chen}
\affiliation{Jockey Club Institute for Advanced Study, The Hong Kong University of Science and Technology, Clear Water Bay, Kowloon, Hong Kong, China}

\author{Xiao-Han Ma}
\affiliation{Deep Space Exploration Laboratory/School of Physical Sciences, University of Science and Technology of China, Hefei, Anhui 230026, China}
\affiliation{CAS Key Laboratory for Researches in Galaxies and Cosmology/Department of Astronomy, School of Astronomy and Space Science, University of Science and Technology of China, Hefei, Anhui 230026, China}

\author{Bo Wang}
\email{ymwangbo@ustc.edu.cn}
\affiliation{Deep Space Exploration Laboratory/School of Physical Sciences, University of Science and Technology of China, Hefei, Anhui 230026, China}
\affiliation{CAS Key Laboratory for Researches in Galaxies and Cosmology/Department of Astronomy, School of Astronomy and Space Science, University of Science and Technology of China, Hefei, Anhui 230026, China}

\author{Wentao Luo}
\affiliation{Deep Space Exploration Laboratory/School of Physical Sciences, University of Science and Technology of China, Hefei, Anhui 230026, China}
\affiliation{CAS Key Laboratory for Researches in Galaxies and Cosmology/Department of Astronomy, School of Astronomy and Space Science, University of Science and Technology of China, Hefei, Anhui 230026, China}

\author{Damien A. Easson}
\email{easson@asu.edu}
\affiliation{Department of Physics, Arizona State University, Tempe, AZ 85287-1504, USA}

\author{Antonino Marcian\`o}
\email{marciano@fudan.edu.cn}
\affiliation{Department of Physics \& Center for Field Theory and Particle Physics, Fudan University, 200433 Shanghai, China}
\affiliation{Laboratori Nazionali di Frascati INFN Via Enrico Fermi 54, Frascati (Roma), Italy}
\affiliation{INFN sezione Roma Tor Vergata, I-00133 Rome, Italy}

\begin{abstract}
\noindent
We discuss a novel window to probe the origin of our universe via the mass functions of primordial black holes (PBHs). The mass functions of PBHs are simply estimated using the conventional Press-Schechter formalism for two paradigms of cosmic origin, including inflationary $\Lambda$CDM and bounce cosmology. The standard inflationary $\Lambda$CDM model cannot generate an appreciable number of massive PBHs; however, non-trivial inflation models with blue-tilted power spectra at small scales and matter bounce cosmology provide formation mechanisms for heavy PBHs, which in turn, may seed the observed supermassive black holes (SMBHs). By fitting the SMBH mass functions at high redshift ($z \sim 6$) derived from Sloan Digital Sky Survey (SDSS) and Canada-France High-z Quasar Survey (CFHQS) quasars, for two paradigms of cosmic origin, we derive constraints on the PBH density fraction $f_{\mathrm{PBH}}$ at $z \sim 6$ and the characteristic mass $M_{\star}$, with the prior assumption that all SMBHs stem from PBHs. We demonstrate that this newly proposed procedure, relying on astronomical measurements that utilize deep-field surveys of SMBHs at high redshift, can be used to constrain models of cosmic origin. Additionally,  although not the main focus of this paper, we evolve the mass function from $z\sim6$ to $z\sim0$ through an assumption of $3\times 10^8$-year Eddington's accretion, and give a rough estimation of $f_{\mathrm{PBH}}$ at $z \sim 0$.
\end{abstract}


\maketitle
\section{Introduction}

How to probe cosmic origin is one of the most fundamental questions in modern physics, and sets the cornerstone supporting modern techniques for experimental development and  for observational cosmology. Over decades, we have precisely measured a nearly scale-invariant power spectrum of primordial density fluctuations through the windows provided by the cosmic microwave background (CMB) \cite{Planck18_5PowerSpectrum} and large-scale structure (LSS) \cite{eBOSS:2020yzd, DES:2021zxv}. The measurements yield significant constraints on the inflationary $\Lambda$CDM model, first proposed to understand initial conditions for hot big bang cosmology \cite{Guth:1980zm, Starobinsky:1980te, Linde:1981mu, Albrecht:1982wi}. However, research interests in alternative scenarios remain extensive, both to attain developments of cosmological perturbation theory \cite{Mukhanov:1990me} and to face challenges from several astronomical experiments \cite{Bull:2015stt, Bullock:2017xww, Verde:2019ivm, DiValentino:2021izs}. Cosmologists are motivated to search for primordial paradigms beyond inflation, including bounce cosmology \cite{Brandenberger:1993ef, Peter:2008qz, Cai:2012va, Cai:2014bea, Brandenberger:2016vhg}, ekpyrotic cosmology \cite{Khoury:2001wf, Steinhardt:2001st}, string gas cosmology \cite{Brandenberger:1988aj, Nayeri:2005ck, Battefeld:2005av} and emergent universe \cite{Ellis:2002we, Ellis:2003qz, Creminelli:2010ba, ilyas2021emergent}, among others. Naturally, the question follows whether or not these theoretical paradigms can be distinguished by observational physics.

While many efforts have been made on the precise characterization of power spectra of primordial perturbations on large scales \cite{Brandenberger:2006pr, cai2016bouncing}, it becomes crucial to look for new and independent probes at small scales. A novel prediction of primordial perturbations at small scales is the formation of primordial black holes (PBHs) \cite{1967SvA....10..602Z, Hawking:1971ei, Carr:1974nx, Sasaki:2018dmp, Khlopov:2008qy,Wang:2022nml, Carr:2018rid}. Due to their broad mass distribution, PBHs can possibly seed supermassive black holes (SMBHs) which were found in abundance in astronomical surveys \cite{Volonteri:2010wz, 1989IAUS..134..217D, 2009ApJ...694..302D}. In particular, many SMBHs were discovered at high redshifts while the underlying formation mechanism remains unclear \cite{Magorrian:1997hw, gultekin2009m, McConnell:2011mu}.
The current astrophysical approaches of SMBH formation suffer puzzles such as supercritical accretion time, exceptionally high star formation rate, etc \cite{2004ApJ...613...36H, Shapiro:2004ud, Volonteri:2010wz, Inayoshi:2019fun}. Accordingly, besides their astrophysical origin, it is necessary to consider the cosmological origin of SMBHs formed in the primordial epoch. It is well known that a nearly scale-invariant power spectrum derived from inflationary $\Lambda$CDM cosmology can hardly generate enough PBHs responsible for the observed SMBHs \cite{Bean:2002kx}. Thus, several paradigms of cosmic origin that may generate sufficient PBHs were extensively pursued, such as non-trivial inflation models with a blue-tilted power spectrum at small scales \cite{Byrnes:2018txb, carr2020primordial, 2021RPPh...84k6902C} and bounce cosmology \cite{Quintin:2016qro, chen2017tracing, Chen:2022usd, Carr:2011hv, Carr:2017wkz}.

In this paper, we introduce a novel probe of cosmic origin using the mass function of PBHs predicted by different cosmological scenarios and comparing it with the observed SMBHs. Specifically, we consider the inflationary scenario that allows a blue-tilted power spectrum at small scales, dubbed as small-scale enhanced inflation (SSE-inflation), and the cosmology of a matter bounce in which the regular thermal expansion is preceded by a period of matter-dominated contraction. This approach may be applied to other cosmic origin models as well. For both scenarios, we consider power-law-parameterized primordial power spectra. We then adopt the conventional Press-Schechter (PS) formalism \cite{press1974formation, Carr:1975qj} to estimate their PBH mass functions under the assumption of Gaussian probability distributions of density perturbations. 
As a proof-of-concept, we consider the simplest case that all SMBHs are generated from PBHs.
Utilizing the best-fit SMBH mass function at $z\sim 6$ from Ref.~\cite{Willott:2010yu}, we perform Markov Chain Monte Carlo (MCMC) simulations to pin down the parameter spaces of the fraction of PBHs against total dark matter $f_{\mathrm{PBH}}(6)$ at $z \sim 6$ and the characteristic mass $M_{\star}$ for different cosmological scenarios. Our analysis involves a comparison of the mass functions of massive PBHs produced within distinct scenarios with the observed SMBHs. Notably, we find that the SSE-inflation model exhibits a superior fit to the observational data in contrast to the matter bounce model.
Additionally, taking into account Eddington's accretion over a period of $3 \times 10^8$ years from $z\sim 6$ to $z\sim 0$, we estimate $f_{\mathrm{PBH}}(0)$ at $z\sim 0$ for the SSE-inflation model and compare the resulting mass function with observational data at $z\sim0$ \cite{marconi2004local,vika2009millennium,li2011cosmological,2020NatAs...4..282S,sicilia2022black}. As a result, investigating the feature of supermassive PBHs mass function incorporated with the primordial density power spectra, provides a novel probe to falsify cosmic origin paradigms by observing SMBHs.

\section{ Power spectra and PBH formation}

Several cosmological surveys, including CMB, LSS, galaxy clustering, Lyman-$\alpha$ forest, etc., have tightly constrained the matter power spectrum at scales $k \lesssim 3{\rm ~Mpc}^{-1}$ \cite{reid2010cosmological, chabanier2019matter}. These observations are consistent with the predictions of the standard inflationary $\Lambda$CDM model, according to which the primordial power spectrum $P(k) = A_s (k/k_*)^{n_s-1}$ is nearly scale-invariant, with spectral index $n_s$ and amplitude $A_s$ measured precisely at pivot scale $k_*$ by Planck \cite{Planck18_6CosmicPara}. This red-tilted power spectrum prevents PBHs from constituting a considerable fraction of dark matter \cite{carr2020primordial,2021RPPh...84k6902C}, seeding SMBHs \cite{Bean:2002kx}, and from addressing several such issues as well. Theoretically, however, there are many realizations of SSE-inflation that generate a blue-tilted spectrum at small scales, allowing for an increase in the abundance of massive PBHs, such as non-attractor inflation \cite{Kinney:2005vj, Martin:2012pe, Garcia-Bellido:2017mdw} and non-perturbative resonance effects \cite{Cai:2018tuh, Cai:2019jah, Zhou:2020kkf, Cai:2021yvq}. Furthermore, it was found in \cite{Quintin:2016qro, chen2017tracing} that a matter bounce can produce substantial PBH formation.

To systematically investigate the formation of PBHs, we adopt the general forms of power spectra for SSE-inflation and matter bounce \cite{sureda2021press, 2011JCAP...03..003C}, which are parameterized in the power-law form by,
\begin{equation} \label{Eq:PowerSpectrum}
\begin{aligned}
 &P_{\mathcal{R}}^I(k) = \begin{cases}A_s {(k/k_*)}^{n_s-1} ~,~ k<k_{t} \\
 A_b{(k/k_*)}^{n_{b}-1} ~,~ k \geqslant k_{t}\end{cases} ~;
 \\
 &P_{\mathcal{R}}^B(k, a) \simeq A_B k^{n_B-1} a^{-3} ~,
\end{aligned}
\end{equation}
where ``$I$" and ``$B$" denote the SSE-inflation and matter bounce scenarios, respectively. For SSE-inflation, a characteristic scale $k_{t}$ is introduced to account for the transition from the red-tilted spectrum with amplitude $A_s$ and spectral index $n_s \sim 1$ to the blue-tilted spectrum with $A_b$ and $n_b >1$, the latter induces an efficient formation of heavy PBHs.

Note that, the parametrizations of the blue-tilted spectra \eqref{Eq:PowerSpectrum} are not necessarily applicable at smaller scales, due to the uncertainty of the non-linear effects (inside the galactic scales). Nonetheless, these small-scale density perturbations have negligible effect on the formation of supermassive PBHs which are determined by the relatively large-scale density perturbations.

For matter bounce, the power spectrum is parameterized by the corresponding amplitude $A_B$ and spectral index $n_B$, and inversely proportional to the cubic power of the scale factor $a$. During the contracting phase, the scalar perturbations grow as the universe contracts \cite{2011JCAP...03..003C, Peter:2006hx}.

In addition to the distinct shapes of the primordial power spectra, the formation mechanisms of PBHs are also different in the scenarios of SSE-inflation and matter bounce. For SSE-inflation, PBHs can form when the amplified perturbations re-enter the Hubble radius during radiation domination \cite{Carr:1974nx}; while for matter bounce, when the Jeans length is smaller than the Hubble radius during the contracting phase, quantum fluctuations at the Jeans-length crossing can evolve into classical perturbations at super-Jeans, but sub-Hubble scales. Thus, the growth of perturbations in the super-Jeans/sub-Hubble regime can allow for the formation of PBHs \cite{Carr:2011hv, Quintin:2016qro, chen2017tracing}. Consequently, the PBH mass functions derived in these two scenarios are expected to exhibit different behaviors.

\section{PBH initial mass functions}

The initial PBH mass function when accretion just starts is derived by the PS formalism \cite{Carr:2016drx, sureda2021press},
\begin{equation}
\label{Eq:MassFunction}
 \left(\frac{\mathrm{d} n}{\mathrm{d} M}\right)^I=\frac{ 2 A_{n}f^I_{\mathrm{PBH}}\rho^I_{\mathrm{DM}}}{\sqrt{2 \pi} M} \Big| \frac{\mathrm{d} \nu}{\mathrm{d} M} \Big| e^{-\frac{1}{2}\nu(M)^2} ~,
\end{equation}
where $f^I_{\mathrm{PBH}}$ is the PBH fraction against total dark matter at PBH formation, $\rho^I_{\mathrm{DM}}$ is the dark matter density at PBH formation,
$A_n \sim  1$ is a normalized factor \cite{sureda2021press} and $\nu(M)\equiv \delta_c/\sigma(M)$ is the peak height with critical density contrast $\delta_c$ and fluctuation variance $\sigma(M)$. 
It is manifest from Eq.~\eqref{Eq:MassFunction} that the shape of the PBH mass function relies on $\nu(M)$, hence indicates an exponential suppression of the PBH mass function, controlled by the characteristic mass $M_{\star}$ defined as $\nu(M_{\star})=1$. Using the power spectra presented in Eq.~\eqref{Eq:PowerSpectrum}, it is straightforward to calculate the PBH mass functions for SSE-inflation and matter bounce via the formula \eqref{Eq:MassFunction} through the following peak heights \cite{sureda2021press,Quintin:2016qro}:
\begin{equation} \label{Eq:PeakHeight}
\begin{aligned}
 &\nu_{I}(M) = \Big( \frac{S_1 M_{\star}^2+S_2 M_{\star}^{-\beta}}{S_1 M^2+S_2 M^{-\beta}} \Big)^{1/2} ~;\\
 &\nu_{B}(M) = \Big( \frac{M}{M_{\star}} \Big)^{\frac{n_B+5}{6}} ~,
\end{aligned}
\end{equation}
where $\beta=(n_b-1)/2$, $S_1=(n_b-n_s) / M_{t}^{2+\beta}$ and $S_2=n_s+3$, with a particular mass $M_{t}\sim  4\pi^2 H_0^2 \Omega_{r,0}/k_{t}^2$, $H_0$ denoting the present Hubble constant and $\Omega_{r,0}$ the radiation density parameter. Distinct forms of $\nu(M)$ within different scenarios allow one to distinguish the related mass functions. This is triggered by calculations of $\sigma(M)$ through the PS formalism, which depends both on the power spectrum of density perturbation and the window function. Eq.~\eqref{Eq:PowerSpectrum} leads to different density perturbation power spectra. In addition, the scales at which PBH formation occurs differently in the expanding and contracting phases and hence affect the smoothing scale of the window function.

For SSE-inflation, $\nu_I(M)$ increases with $M$, reaches the maximum at $M\sim \mathcal{O} (M_t)$, and then starts to decrease with larger mass $M$. Accordingly, the mass function vanishes at the maximum of $\nu_I(M)$, while for masses larger than $\mathcal{O} (M_t)$, its value is controlled by the red-tilted spectrum at the scales $k<k_t$. However, as the efficiency of the PBH production at large scales is suppressed due to a small amplitude of power spectrum constrained by CMB \cite{Green:2020jor}, it is reasonable to introduce a cut-off for the PBH mass function at $M \sim \mathcal{O}(M_t)$ without loss of accuracy in SSE-inflation.
A relatively large $M_t$ is then required to be consistent with the mass of produced PBHs, as seeds of observed SMBHs. The lower bound of $k_t$ is on the other side arising from the observed nearly scale-invariant spectrum at large scales. Numerical results show that $k_t \sim \mathcal{O}(10)~ \mathrm{Mpc}^{-1}$ and that the PBH mass function is insensitive to $k_t$. Moreover, the slopes of the PBH mass functions ($M < M_\star$) can be estimated from Eq.~\eqref{Eq:PeakHeight}, namely $\mathrm{d} n/\mathrm{d} M \propto M^{\frac{n_b-9}{4}}$ and $\propto M^{\frac{n_B-7}{6}}$ for SSE-inflation and matter bounce, respectively.

\begin{figure}[htp!]
\centering
\includegraphics[width=0.95\columnwidth]{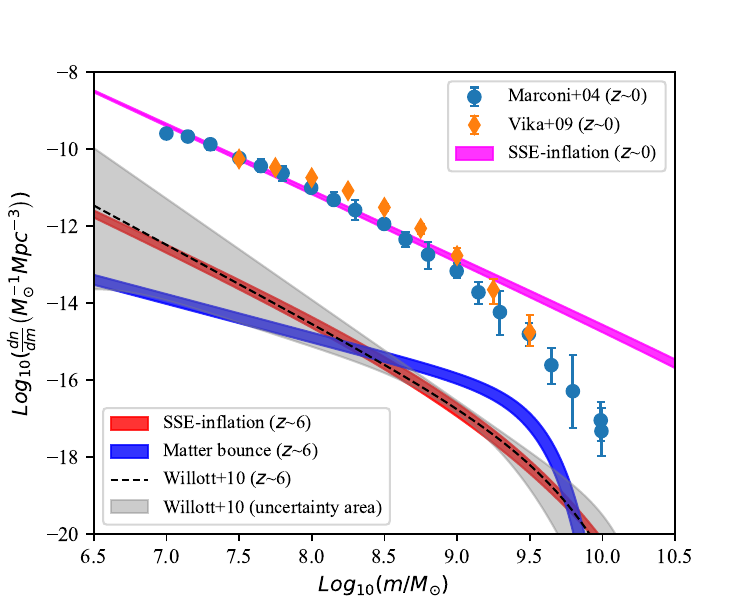}
\caption{PBH mass functions derived from the best-fit parameters at $68\%$ C.L. in SSE-inflation (red shaded region) and matter bounce  (blue shaded region), by using observational SMBH mass functions at $z\sim 6$ from Willott {\it et al.} \cite{Willott:2010yu} (black dashed line). The gray shaded region depicts the uncertainty region derived from bootstrap resampled mass functions \cite{Willott:2010yu}. Magenta shaded region shows the theoretical mass function at $z\sim0$ evolved from the result of SSE-inflation at $z\sim6$. As a comparison, observational SMBH mass functions at $z\sim0$ from Marconi {\it et al.} \cite{marconi2004local} (blue circles), Vika {\it et al.} \cite{vika2009millennium} (orange diamonds) are shown.}
\label{Fig:MassFunction}
\end{figure}
\section{Late-time evolution}

Taking into account the dynamical processes in the late-time evolution of PBHs, for example, accretion and merger, one can rewrite the PBH mass function Eq.~\eqref {Eq:MassFunction} at a specific redshift $z$ as:
\begin{equation}
\label{Eq:MassFunction1}
 \frac{\mathrm{d} n}{\mathrm{d} m}(m,z)=\frac{ f_{\mathrm{PBH}}(z)\rho_{\mathrm{DM}}(z)}{f^I_{\mathrm{PBH}}\rho^I_{\mathrm{DM}}}  \frac{\mathrm{d}M}{\mathrm{d} m}\left(\frac{\mathrm{d} n}{\mathrm{d} M}\right)^I ~,
\end{equation}
where $z$ denotes the redshift of each parameter,
and $m(M)$ is the PBH mass at redshift $z$ which evolved from $M$. 
It is evident from Eq.~\eqref{Eq:MassFunction1} that the evolution of the PBH mass function is highly model-dependent. For the sake of simplicity and clarity, we adopt one accretion model from Ref.~\cite{lapi2014coevolution}. In this model, the SMBHs experience a super-Eddington accretion with growth time $t=3 \times 10^8$ years before $z\simeq 6$ with an effective time-scale $\tau_{eff}=1.8 \times 10^7$ years. It leads to a growth of mass as $\frac{m(z\sim 6)}{M} \simeq  e^{t/\tau_{eff}}\simeq 10^{7.238}$. Hence, by utilizing Eq.~\eqref{Eq:MassFunction1}, we derive the mass function at $z\simeq 6$ for subsequent numerical calculations in the following section.

\section{Numerical results}

In the following numerical calculation, we choose the best-fit SMBH mass functions at $z\sim 6$ obtained from the analysis of 40 quasars in the SDSS and CFHQS \cite{Willott:2010yu}. We perform the MCMC simulations ({\tt{emcee}} \cite{Foreman-Mackey:2012any}), exploiting the observational SMBH mass functions provided in Ref.~\cite{Willott:2010yu}, and derive the best-fit model parameters $M_{\star}$, $f_{\mathrm{PBH}} (6)$ and $n_b$. The PBH mass functions for SSE-inflation and matter bounce, derived from the best-fit values at $68\%$ confidence level (C.L.), are shown by the red and the blue shaded regions in Fig.~\ref{Fig:MassFunction}, respectively. We set the transition scale to be $k_{t} = 10 ~ \text{Mpc} ^{-1}$, which corresponds to the mass $M_{t}= 9\times 10^{10} M_{\odot}$ to ensure that the mass of produced PBHs is consistent with the mass range of SMBHs.
Without loss of generality, in SSE-inflation, we choose the spectral index of the blue-tilted power spectrum $n_b$ to be a free parameter while keeping the power spectrum in the matter bounce to be almost scale-invariant, i.e., $n_B\sim1$ in order to be consistent with cosmological observations \cite{Quintin:2016qro, Peter:2006hx}.

From Fig. \ref{Fig:MassFunction}, we observe that the mass function in the SSE-inflation scenario fits the observation  at $z\sim 6$ better than the matter bounce scenario. Note that there are two significant behaviors.
First, within the lower-mass region ($m \lesssim 10^9M_{\odot}$), the slopes of mass functions for both scenarios, derived through the numerical calculations, are completely consistent with the analytical estimates in the previous section.
Second, for the larger-mass region ($m \gtrsim  10^9M_{\odot}$), the matter bounce yields a mass function that decreases faster than that of SSE-inflation. The reason is that the monotonic growth of $\nu_B$ leads to a drastic exponential suppression of the mass function in the matter bounce scenario. 
Therefore, these crucial characters and more precise SMBH surveys can be exploited in order to distinguish different cosmic origin paradigms in the near future.

\begin{figure}[htp!]
\centering
\includegraphics[width=0.95\columnwidth]{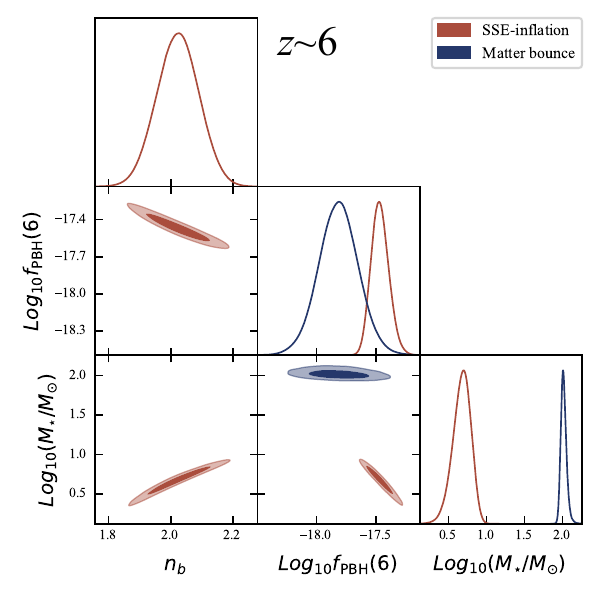}
\caption{The best-fit values of the parameters $M_\star$, $f_\mathrm{PBH}(6)$ and $n_b$ from MCMC simulations of SSE-inflation (red) and matter bounce (blue) scenarios by fitting the data at $z\sim 6$ \cite{Willott:2010yu}. In each scenario, the $68\%$ C.L. (deep color) and $95\%$ C.L. (light color) regions of each parameter are provided.}
\label{Fig:MCMCResult1}
\end{figure}
Fig.~\ref{Fig:MCMCResult1} illustrates the parameter space for $M_{\star}$ and $f_{\mathrm{PBH}} (6)$ within $1 \sigma$- and $2 \sigma$-confidence regions for both the SSE-inflation and matter bounce scenarios, respectively.
The best-fit parameter space for SSE-inflation at $68\%$ C.L. reads
\begin{equation}
\begin{aligned}
& \displaystyle{\mathrm{Log}_{10}\Big(\frac{M_\star}{M_\odot}\Big) = 0.682_{-0.128}^{+0.110}} ~, \\ 
&\displaystyle{\mathrm{Log} _{10}f_{\mathrm{PBH}} (6) = -17.470_{-0.071}^{+0.077}} ~,
\\&
n_b=2.023 _{-0.068}^{+0.067} ~, \nonumber
\end{aligned}
\end{equation}
while for matter bounce at $68\%$ C.L. reads
\begin{equation}
\begin{aligned}
&\displaystyle{\mathrm{Log}_{10}\Big(\frac{M_\star}{M_\odot}\Big) = 2.013_{-0.033}^{+0.040}} ~,\\
&\displaystyle{\mathrm{Log} _{10}f_{\mathrm{PBH}} (6) = -17.814 _{-0.166}^{+0.164}} ~. \nonumber
\end{aligned}
\end{equation}
For comparison, we perform the simulations with the condition that processes like accretion or merger are omitted before $z\sim 6$. The results indicate that in the absence of these processes, $M_{\star} \sim 10^8 M_{\odot}$ and $f_{\mathrm{PBH}} (6)\sim 10^{-10}$. This reveals a significant impact from the accretion effect, which notably narrows down the parameter space.
We also list the observational SMBH mass function at $z\sim0$ from SDSS \cite{marconi2004local} and Millennium Galaxy Catalogue survey \cite{vika2009millennium} in Fig.~\ref{Fig:MassFunction}. Interestingly, if we adopt the conventional Eddington's accretion with the duration $t_{\rm grow}\simeq 3 \times 10^8 $ years \cite{lapi2014coevolution} and the accretion rate $\frac{m(0)}{m(6)} \simeq  e^{t_{\rm grow}/t_{\mathrm{ED}}}\simeq 10^{3.257}$ (where $t_{\mathrm{ED}} \approx 4 \times 10^7$ years is the Salpeter time-scale \cite{Carr:2018rid}), the mass function at $z \sim 0$ predicted by SSE-inflation (the magenta region in Fig.~\ref{Fig:MassFunction}) can be also derived from the corresponding mass function at $z\sim6$ (the red region in Fig.~\ref{Fig:MassFunction}) with such accretion process. 
The predicted mass function fits the observational data for $m<10^9 M_\odot$, but deviates from observation as $m$ increases. 
Note that this is a rough estimate without taking into account the merger process of SMBHs or additional effects which are beyond the scope of this paper.
The current PBH fraction $f_\mathrm{PBH}(0)$ for the SSE-inflation with best fit parameters can be estimated as  \cite{PhysRevD.102.043505}
\begin{equation}\label{fpbh0}
f_\mathrm{PBH}(0)
\simeq\frac{ m(0)}{m(6)\left[f_{\mathrm{PBH}}^{-1}(6)-1\right]+m(0)}
\simeq10^{-14}  ,\nonumber
\end{equation}
which is consistent with the current observational constraints from $\mu$-distortion $f_\mathrm{PBH}(0)<10^{-5}$ \cite{ricotti2008effect,Fixsen:1996nj}.

According to the numerical results, the spectral index $n_b$ indicates a blue-tilted spectrum, so that the PBH produced by the enhanced power spectrum at small scales can account for SMBHs in the SSE-inflation. This is not the case for the matter bounce, for which the power spectrum remains nearly scale-invariant at all scales. However, the abundance of PBHs can be realized by increasing the growth history of the contracting phase.
Consequently, these two scenarios can be further distinguished using the measurements of the scalar power spectrum at $10~\mathrm{Mpc}^{-1} \lesssim k \lesssim 10^9~ \mathrm{Mpc}^{-1}$, by exploiting data from forthcoming surveys, e.g., Primordial Inflation Explorer \cite{Kogut:2011xw}, Square Kilometre Array \cite{Moore:2014lga,2009IEEEP..97.1482D}.

\section{Concluding remarks}

We proposed a promising scheme that offers the possibility to probe the origin of the universe by combining PBH mass functions with the observed SMBHs. Every cosmic origin paradigm predicts a unique scalar power spectrum at small scales and thus leads to distinct PBH mass functions. Specifically, we considered two representative scenarios, SSE-inflation and the matter bounce. For the sake of generality, we parameterized the primordial curvature spectra in the power-law form and adopted the conventional PS formalism to estimate the resulting PBH initial mass functions. 
By utilizing the accretion model from Ref.~\cite{lapi2014coevolution}, we evolve the initial mass functions to mass functions at redshifts $z\sim 6$.
Incorporating observational SMBH mass functions at redshifts $z\sim 6$, by performing the MCMC numerical computation, we found the best-fit values of the parameters $M_\star$, $f_\mathrm{PBH}(6)$, and $n_b$. The results show that the behavior of the mass functions for different cosmological scenarios in the mass range of SMBHs is distinguishable. And the mass function in SSE-inflation fits the observation better. Subsequently, we track the evolution of the mass function within the SSE-inflation model from $z\sim6$ to $z\sim0$, which is achieved through the consideration of a simple Eddington’s accretion model. 
The in-depth analysis of astrophysical dynamical processes, such as accretion and merger, should be reserved for future work.

To comprehensively evaluate the significance of our results, it is crucial to recognize that the observational SMBHs at $z \sim 6$ which we used in the numerical analysis is constrained to the mass range $10^8 M_{\odot} - 3 \times 10^9 M_{\odot}$. Recent observations from the James Webb Space Telescope (JWST) and the Near Infrared Spectrograph (NIRSpec) have identified several quasars at $z>6$ \cite{2023ApJ...953L..29L,2022ApJ...940L...1W,2021ApJ...923..262Y}, but the sample size remains insufficient for a detailed mass function analysis. Consequently, our model necessitates additional observational data for more robust validation in the future.

Inspired by the new probe developed in this paper, there are several follow-up issues that deserve further investigation, namely the generalization of the PS formalism to alternative paradigms, including ekpyrotic cosmology and the emergent universe, massive PBH formation in hybrid cosmic origin scenarios, the evolution of resulting mass function throughout the cosmic history, and so forth. Additionally, non-Gaussian effects can significantly affect the PBH formation \cite{Cai:2021zsp, Cai:2022erk, Panagopoulos:2019ail, Ezquiaga:2019ftu, Figueroa:2020jkf, Ferrante:2022mui, Pi:2022ysn,Nakama:2017xvq}, and may also lead to distinguishable features for this probe. These topics bridge the theoretical study of cosmic origin and observations of SMBHs, enriching our approach toward the investigation of the early universe.

We conclude with the statement that this novel probe, along with the accumulated data of SMBHs, is expected to shed light on cosmic origin. Remarkably, a population of SMBHs at high redshift has been found by several observations, such as SDSS, Chandra X-ray Observatory, and DESI Legacy Imaging Surveys, etc \cite{SDSS:2001emm, Wang:2020oly, yang2020poniua}. The current James Webb Space Telescope also revealed the discovery of extremely high luminosity objects that could be SMBH candidates at high redshift \cite{Wylezalek:2022nts}. Those observations indicate possible challenges to the theory of primordial structure formation within the standard $\Lambda$CDM cosmology. According to our results, for the alternative scenarios, the structure formations are allowed at much earlier epochs. It becomes then experimentally possible to falsify several cosmic origin scenarios through forthcoming astronomical observations derived by means of deep-field scans of SMBHs at high redshifts.

\section{Acknowledgments}

We thank anonymous referees for their insightful comments and suggestions. We are also grateful to Bernard Carr, Yun Fang, Ricardo Z. Ferreira, Amara Ilyas, Ugo Moschella, Jerome Quintin, Misao Sasaki, Dong-Gang Wang, Yi Wang, Zihan Zhou, and Mian Zhu for valuable communications. This work is supported in part by National Key R\&D Program of China (2021YFC2203100), by CAS Young Interdisciplinary Innovation Team (JCTD-2022-20), by NSFC (11875113, 11961131007, 12261131497, 12003029, 11833005, 12192224), by 111 Project for "Observational and Theoretical Research on Dark Matter and Dark Energy" (B23042), by Fundamental Research Funds for Central Universities, by CSC Innovation Talent Funds, by USTC Fellowship for International Cooperation, by USTC Research Funds of the Double First-Class Initiative, by CAS project for young scientists in basic research (YSBR-006), by the Shanghai Municipality Science and Technology Commission (KBH1512299). SFY is supported by the Disposizione del Presidente INFN n. 24433 in INFN Sezione di Milano. DE is supported in part by the U.S. Department of Energy, Office of High Energy Physics, under Award No. DE-SC0019470, and the Foundational Questions Institute under Grant No. FQXi-MGB-1927. We acknowledge the use of the computing cluster {\it LINDA} \& {\it JUDY} in the particle cosmology group at USTC.

\bibliography{pbhletter}

\end{document}